\documentclass[a4paper,11pt]{article}
\usepackage{lmodern}

\usepackage[T1]{fontenc}
\usepackage[latin9]{inputenc}
\usepackage{amsmath}
\usepackage{graphicx}

\makeatletter
\newcommand{\lyxaddress}[1]{
	\par {\raggedright #1
	\vspace{1.4em}
	\noindent\par}
}

\usepackage{mathrsfs}   
\usepackage{slashed}     
\usepackage{bbold}  
\usepackage{url}
\usepackage{graphicx}
\usepackage[colorlinks=true,linkcolor=redLinks,citecolor=greenLinks,urlcolor=redLinks, pdfborder={0 0 1}]{hyperref}
\usepackage{xcolor}
\usepackage{framed}
\usepackage[numbers,sort&compress]{natbib}

\colorlet{shadecolor}{gray!15}

\definecolor{greenLinks}{rgb}{0, 0.6, 0} 
\definecolor{blueLinks}{rgb}{0, 0, 0.6}
\definecolor{redLinks}{rgb}{0.6, 0, 0}
\definecolor{tempText}{rgb}{0.55, 0.10,0.67}
\definecolor{eprintLinks}{rgb}{0.4, 0.4, 0.4}
\definecolor{journalLinks}{rgb}{0.6, 0, 0}

\newcommand{\MYhref}[3][redLinks]{\href{#2}{\color{#1}{#3}}}%

\usepackage{multirow}
\let\orig@Hy@EveryPageAnchor\Hy@EveryPageAnchor
\def\Hy@EveryPageAnchor{%
    \begingroup
    \hypersetup{pdfview=Fit}%
    \orig@Hy@EveryPageAnchor
    \endgroup
}

\let\oldFootnote\footnote
\newcommand\nextToken\relax

\renewcommand\footnote[1]{%
    \oldFootnote{#1}\futurelet\nextToken\isFootnote}

\newcommand\isFootnote{%
    \ifx\footnote\nextToken\textsuperscript{,}\fi}

\definecolor{myPurple}{RGB}{128,0,182}

\makeatother

\begin{document}
\title{Boundedness from below of $SU(n)$ potentials}
\author{Renato M. Fonseca\date{}}
\maketitle

\lyxaddress{\begin{center}
{\Large{}\vspace{-0.5cm}}High Energy Physics Group\\
Departamento de Física Teórica y del Cosmos,\\
Universidad de Granada, E--18071 Granada, Spain\\
~\\
Institute of Particle and Nuclear Physics\\
Faculty of Mathematics and Physics, Charles University,\\
V Holešovi\v{c}kách 2, 18000 Prague 8, Czech Republic\\
~\\
Email: renatofonseca@ugr.es
\par\end{center}}
\begin{abstract}
Vacuum stability requires that the  scalar potential is bounded from
below. Whether or not this is true depends on the scalar quartic interactions
alone, but even so the analysis is arduous and has only been carried
out for a limited set of models. Complementing the existing literature,
this work contains the necessary and sufficient conditions for two
$SU(n)$ invariant potentials to be bounded from below. In particular,
expressions are given for models with the fundamental and the 2-index
(anti)symmetric representations of this group. A sufficient condition
for vacuum stability is also provided for models with the fundamental
and the adjoint representations. Finally, some considerations are
made concerning the model with the gauge group $SU(2)$ and the scalar
representations $\boldsymbol{1}$, $\boldsymbol{2}$ and $\boldsymbol{3}$;
such a setup is particularly important for neutrino mass generation
and lepton number violation.
\end{abstract}

\section{Introduction}

The study of scalar potentials can be a formidable task given that
these are quartic functions of several variables. Despite the difficulty,
their analysis is crucial as the scalar minima correspond to the possible
vacuum configurations.

A given vacuum state cannot be absolutely stable if the scalar potential
acquires lower values for some other choice of field values. Of particular
concern are those cases where the potential is not bounded from below
(BFB), meaning that it acquires arbitrarily large negative values.
If this were to happened it would be for field values far from the
origin, in which case quadratic and trilinear interactions can be
neglected. Even so, deriving the BFB conditions quickly becomes a
very complicated problem as the number of scalar fields increases,
so much so that in the literature one can find the derivation of these
conditions for just a few models. Among the cases which were considered
there is the two Higgs doublet model \cite{Ivanov:2006yq}, the type-II
seesaw potential with the Higgs doublet plus an $SU(2)$ triplet \cite{Arhrib:2011uy,Bonilla:2015eha},
special three Higgs doublet models \cite{Buskin:2021eig,Faro:2019vcd,Ivanov:2020jra}
and also an $SU(3)$ invariant potential with three triplets \cite{Costantini:2020xrn}.
Several other works have analyzed the vacuum stability of specific
models or discussed general techniques for doing so \cite{Kannike:2012pe,Kannike:2016fmd,Ivanov:2018jmz,Moultaka:2020dmb}.

Of particular relevance to the following discussion is the analysis
in reference \cite{Bonilla:2015eha} on the BFB conditions for the
Standard Model potential with the inclusion of a scalar triplet, which
refined the results in \cite{Arhrib:2011uy}. This corresponds to
an $SU(2)$ invariant potential with the scalar representations $\boldsymbol{2}$
and $\boldsymbol{3}$. Following up on that analysis, the aim of
the present work is threefold:
\begin{enumerate}
\item Generalize the results of \cite{Bonilla:2015eha} to $SU(n)$ invariant
potentials with the fundamental representation plus a 2-index representation
--- the symmetric, the antisymmetric or the adjoint. This last representation
presents a unique difficulty, hence I will only derive a sufficient
condition (which is not a necessary one) for the potential to be bounded
from below.
\item A crucial step in the derivation of the BFB conditions in \cite{Bonilla:2015eha}
--- namely the shape of figure 1 --- was not demonstrated explicitly
up to now, as it was obtained via elaborate manipulations of expressions
in a computer. In this work I provide a fully analytical understanding
of these calculations.
\item The Standard Model potential supplemented by a scalar singlet and
a scalar triplet (a 1-2-3 $SU(2)$ potential, in reference to the
sizes of the irreducible fields) is important in the context of neutrino
mass generation, and also lepton number violation \cite{Schechter:1981cv}.
For such a complicated potential, instead of providing in full generality
the BFB conditions which are both necessary and sufficient, I will
derive them for an important special case where one of the quartic
couplings is neglected. Furthermore, a sufficient condition will be
given for the general case.
\end{enumerate}
It is worth pointing out that extending the results of \cite{Bonilla:2015eha}
to $SU(n)$, with $n>2$, is not a mere mathematical curiosity. Indeed,
it is plausible that the fundamental laws of physics are symmetric
under a group larger than the Standard Model one, such as $SU(3)\times SU(3)\times U(1)$
\cite{Singer:1980sw,Pisano:1992bxx,Frampton:1992wt,Fonseca:2016tbn},
$SU(4)\times SU(2)\times SU(2)$ \cite{Pati:1974yy}, $SU(5)$ \cite{Georgi:1974sy}
and even bigger special unitary groups (see for instance \cite{Fonseca:2015aoa}
and the references contained therein). The viability of the associated
models requires several irreducible scalar representations, in some
cases coinciding with the ones analyzed in this work \cite{Harries:2018tld}.
In other cases, such as the Georgi-Glashow $SU(5)$ model \cite{Georgi:1974sy},
the field content studied in this work is just part of the full scalar
sector, and if so the conditions presented here are still applicable
--- they are necessary (but not sufficient) for the potential to
be bounded from below.

The rest of this document is structured as follows. Section \ref{sec:2}
introduces and analyzes the $SU(n)$ invariant scalar potential with
a fundamental and a 2-index symmetric representation. The BFB conditions
depend on two crucial parameters, $\alpha$ and $\beta$, which are
considered in detail in section \ref{sec:3} plus an appendix. With
a thorough understanding of them, in section \ref{sec:4} I derive
the BFB conditions for the potential mentioned in section \ref{sec:2}
with a 2-index symmetric representation. Some modifications are necessary
in the case of a 2-index anti-symmetric representation, as explained
in section \ref{sec:5}. One can also find there an analysis of the
more complicated setup where the 2-index representation is the adjoint.
The 1-2-3 model mentioned earlier is considered in section \ref{sec:6}.
Finally, for the reader's convenience, a summary of the results can
be found at the very end.

\section{\label{sec:2}An $SU(n)$ invariant potential}

Consider a scalar $\phi_{i}$ transforming under the fundamental representation
of $SU(n)$ as well as a $\Delta_{ij}$ transforming under the 2-index
symmetric representation of this group. These fields can be viewed
as a vector and a matrix which change under an $SU(n)$ transformation
$U$ as follows:
\begin{gather}
\phi\rightarrow U\phi\,,\\
\Delta\rightarrow U\Delta U^{T}\,.\label{eq:Delta-gauge-transformation}
\end{gather}
There are 5 quartic terms allowed by the symmetry, which are
\begin{align}
V^{(4)} & =\frac{\lambda_{\phi}}{2}\left(\phi^{\dagger}\phi\right)^{2}+\frac{\lambda_{\Delta}}{2}\left[\textrm{Tr}\left(\Delta\Delta^{*}\right)\right]^{2}+\frac{\lambda_{\Delta}^{\prime}}{2}\textrm{Tr}\left(\Delta\Delta^{*}\Delta\Delta^{*}\right)+\lambda_{\phi\Delta}\left(\phi^{\dagger}\phi\right)\textrm{Tr}\left(\Delta\Delta^{*}\right)+\lambda_{\phi\Delta}^{\prime}\phi^{\dagger}\Delta\Delta^{*}\phi\,.\label{eq:scalar-potential}
\end{align}
The field $\Delta$ has $n(n+1)/2$ independent components, but it
is always possible to cast $\Delta$ in a diagonal form $\textrm{diag}\left(\Delta_{1},\Delta_{2},\cdots,\Delta_{n}\right)$
with a gauge transformation. In this basis,\footnote{One can also make all $\phi_{i}$ --- or all $\Delta_{i}$ --- real
and non-negative. I will nevertheless abstain from making this further
simplification.} the quartic potential reads
\begin{align}
V^{(4)} & =\frac{\lambda_{\phi}}{2}\left(\sum_{i}\left|\phi_{i}\right|^{2}\right)^{2}+\frac{\lambda_{\Delta}}{2}\left(\sum_{i}\left|\Delta_{i}\right|^{2}\right)^{2}+\frac{\lambda_{\Delta}^{\prime}}{2}\sum_{i}\left|\Delta_{i}\right|^{4}\nonumber \\
 & +\lambda_{\phi\Delta}\left(\sum_{i}\left|\phi_{i}\right|^{2}\right)\left(\sum_{i}\left|\Delta_{i}\right|^{2}\right)+\lambda_{\phi\Delta}^{\prime}\sum_{i}\left|\phi_{i}\right|^{2}\left|\Delta_{i}\right|^{2}\,.
\end{align}
The above expression depends only on the $2n$ non-negative variables
$\left|\phi_{i}\right|^{2}$ and $\left|\Delta_{i}\right|^{2}$, and
the dependence is quadratic. Hence one can in principle use the co-positivity\footnote{A matrix $M$ is co-positive if for every vector $x\neq0$ with real
and non-negative entries it is true that $x^{T}Mx>0$ (sometimes the
sign $\geq$ is considered instead). The fact that the entries of
the vector cannot be negative is crucial. While this might seem a
concept which is too specific to be useful in generic calculations,
its importance and usefulness in the assessment of the stability of
scalar potentials is well established.} conditions \cite{Kannike:2012pe} for a $2n$-dimensional matrix
to infer the values of the $\lambda$ parameters for which $V^{(4)}$
is always positive. The problem is that these conditions become quite
complicated for square matrices with 4 or more rows. I will therefore
follow an approach in line with \cite{Bonilla:2015eha} which is more
readily applicable to variable $n$'s.

Note that with a rescaling
\begin{gather}
\left|\phi_{i}\right|^{2}\rightarrow\frac{1}{\sqrt{\lambda_{\phi}}}\left|\widetilde{\phi}_{i}\right|^{2}\\
\left|\Delta_{i}\right|^{2}\rightarrow\frac{1}{\sqrt{\lambda_{\Delta}+\lambda_{\Delta}^{\prime}}}\left|\widetilde{\Delta}_{i}\right|^{2}
\end{gather}
one can deduce that whether or not the potential is bounded from below
must depend on the 5 $\lambda$'s only through the 3 combinations
\begin{equation}
\kappa_{\Delta}^{\prime}\equiv\frac{\lambda_{\Delta}^{\prime}}{\lambda_{\Delta}+\lambda_{\Delta}^{\prime}}\,,\;\kappa_{\phi\Delta}\equiv\frac{\lambda_{\phi\Delta}}{\sqrt{\lambda_{\phi}}\sqrt{\lambda_{\Delta}+\lambda_{\Delta}^{\prime}}}\,,\;\kappa_{\phi\Delta}^{\prime}\equiv\frac{\lambda_{\phi\Delta}^{\prime}}{\sqrt{\lambda_{\phi}}\sqrt{\lambda_{\Delta}+\lambda_{\Delta}^{\prime}}}\,
\end{equation}
plus the signs of $\lambda_{\phi}$ and $\lambda_{\Delta}+\lambda_{\Delta}^{\prime}$,
which need to be positive. Indeed, to check that this last statement
is true it suffices to consider the specific field directions where
only $\phi_{1}$ is non-zero, and also the case when only $\Delta_{1}$
is non-zero. Despite the allure of working with only 3 $\kappa$'s,
I will not use these them in the following discussion.

Let us now introduce the variables\footnote{I am assuming that at least one $\left|\phi_{i}\right|$ and at least
one $\left|\Delta_{i}\right|$ is non-zero. If $\sum_{i}\left|\Delta_{i}\right|^{2}=0$
then $V^{(4)}$ is positive iff $\lambda_{\phi}>0$ (a condition which
has already been mentioned), while if $\sum_{i}\left|\phi_{i}\right|^{2}=0$
it is required (and sufficient) that $\lambda_{\Delta}+\lambda_{\Delta}^{\prime}>0$
and also $\lambda_{\Delta}+\lambda_{\Delta}^{\prime}/n>0$. This last
condition has not been mentioned in the text yet, but it will appear
eventually, so there is no loss of generality in considering that
$\sum_{i}\left|\phi_{i}\right|^{2},\sum_{i}\left|\Delta_{i}\right|^{2}\neq0$.}
\begin{equation}
\alpha\equiv\frac{\sum_{i}\left|\Delta_{i}\right|^{4}}{\left(\sum_{i}\left|\Delta_{i}\right|^{2}\right)^{2}}\textrm{ and }\beta\equiv\frac{\sum_{i}\left|\Delta_{i}\right|^{2}\left|\phi_{i}\right|^{2}}{\left(\sum_{i}\left|\Delta_{i}\right|^{2}\right)\left(\sum_{i}\left|\phi_{i}\right|^{2}\right)}\label{eq:alpha_beta}
\end{equation}
so that
\begin{align}
V^{(4)} & =\frac{1}{2}\left(\begin{array}{c}
\sum_{i}\left|\phi_{i}\right|^{2}\\
\sum_{i}\left|\Delta_{i}\right|^{2}
\end{array}\right)^{T}\left(\begin{array}{cc}
\lambda_{\phi} & \lambda_{\phi\Delta}+\beta\lambda_{\phi\Delta}^{\prime}\\
\lambda_{\phi\Delta}+\beta\lambda_{\phi\Delta}^{\prime} & \lambda_{\Delta}+\alpha\lambda_{\Delta}^{\prime}
\end{array}\right)\left(\begin{array}{c}
\sum_{i}\left|\phi_{i}\right|^{2}\\
\sum_{i}\left|\Delta_{i}\right|^{2}
\end{array}\right)
\end{align}
This expression is positive if and only if for all values of $\alpha$
and $\beta$ the $2\times2$ matrix above is co-positive.\footnote{The case where all $\left|\phi_{i}\right|^{2}$ and all $\left|\Delta_{i}\right|^{2}$
are simultaneously null is known to lead to $V^{(4)}=0$, therefore
it deserves no further attention.} In turn, that is true if and only if
\begin{gather}
\lambda_{\phi}>0\;\textrm{ and }\lambda_{\Delta}+\alpha\lambda_{\Delta}^{\prime}>0\;\textrm{ and }\lambda_{\phi\Delta}+\beta\lambda_{\phi\Delta}^{\prime}+\sqrt{\lambda_{\phi}\left(\lambda_{\Delta}+\alpha\lambda_{\Delta}^{\prime}\right)}>0\,\label{eq:co-positivity-condition}
\end{gather}
for all values of $\alpha$ and $\beta$. With rather straightforward
steps, we have reduced the initial problem, with $n+n\left(n+1\right)/2$
field directions, first down to $2n$ variables (the $\left|\phi_{i}\right|^{2}$
and the $\left|\Delta_{i}\right|^{2}$) and eventually down to just
two ($\alpha$ and $\beta$). However, to get rid of these remaining
field-dependent quantities, we must first understand what is the range
of values they can take.

\section{\label{sec:3}The allowed values of $\alpha$ and $\beta$}

The price to pay for reducing the $2n$ non-negative field quantities
$\left|\phi_{i}\right|^{2}$ and $\left|\Delta_{i}\right|^{2}$ to
just $\alpha$ and $\beta$ is that the range of the new variable
is not obvious. It is rather easy to see that $\max\left(\alpha\right)=1$
when just one $\left|\Delta_{i}\right|^{2}$ is different from zero,
while on the other hand $\min\left(\alpha\right)=1/n$ is reached
when all $\left|\Delta_{i}\right|^{2}$ have a constant value. As
for $\beta$, if just a single $\phi_{i}$ is different from zero,
and the same is true for the corresponding $\Delta_{i}$ ($\Delta_{j\neq i}=0$)
then we reach a maximum $\beta$ value of 1. If on the other hand
a single $\phi_{i}$ is different from zero and only one $\Delta_{j\neq i}$
is non-null, then $\beta$ reaches a minimum of 0.

So $\alpha\in\left[1/n,1\right]$ and $\beta\in\left[0,1\right]$.
Nevertheless the allowed region for $\left(\alpha,\beta\right)$ is
not a rectangle. For example, when $\alpha$ is minimal ($=1/n$),
all the $\left|\Delta_{i}\right|^{2}$ must have the same value $c$
which means that $\beta$ is forced to be $c\left(\sum_{i}\left|\phi_{i}\right|^{2}\right)/\left[\left(\sum_{i}\left|\phi_{i}\right|^{2}\right)nc\right]=1/n$
as well.

The border of the allowed area for $\left(\alpha,\beta\right)$ can
be found following a generic method proposed long ago in \cite{Frautschi:1981jh,Cummins:1985vg}.
These two quantities can be seen as functions of the variables $\left|\phi_{i}\right|^{2}$
plus the $\left|\Delta_{i}\right|^{2}$, and at the border the vectors
$\left(\partial\alpha/\partial\left|\phi_{i}\right|^{2},\partial\beta/\partial\left|\phi_{i}\right|^{2}\right)^{T}$
and $\left(\partial\alpha/\partial\left|\Delta_{j}\right|^{2},\partial\beta/\partial\left|\Delta_{j}\right|^{2}\right)^{T}$
for all $i$ and $j$ must be proportional to each other (the null
vector $\left(0,0\right)^{T}$ is allowed as well). That is because
at the border of the allowed area for $\left(\alpha,\beta\right)$
it should not be possible to move in two independent directions in
the $\left(\alpha,\beta\right)$ plane by making small variations
of the $\left|\phi_{i}\right|^{2}$ and the $\left|\Delta_{i}\right|^{2}$.
The only caveat is that these last variables cannot be negative, hence
for $\left|\phi_{i}\right|^{2}=0$ and for $\left|\Delta_{i}\right|^{2}=0$
the previous restriction does not apply. Such nuance can be taken
into account by saying that the $2n$ vectors
\begin{equation}
\left|\phi_{j}\right|^{2}\left(\partial\alpha/\partial\left|\phi_{j}\right|^{2},\partial\beta/\partial\left|\phi_{j}\right|^{2}\right)^{T}=x_{j}\left(0,y_{j}-\beta\right)^{T}\label{eq:vector_x}
\end{equation}
and
\begin{equation}
\left|\Delta_{k}\right|^{2}\left(\partial\alpha/\partial\left|\Delta_{k}\right|^{2},\partial\beta/\partial\left|\Delta_{k}\right|^{2}\right)^{T}=y_{k}\left(2\left(y_{k}-\alpha\right),x_{k}-\beta\right)^{T}\label{eq:vector_y}
\end{equation}
must either be null or proportional to some constant vector. The notation
$x_{j}\equiv\left|\phi_{j}\right|^{2}/\sum_{i}\left|\phi_{i}\right|^{2}$
and $y_{j}\equiv\left|\Delta_{j}\right|^{2}/\sum_{i}\left|\Delta_{i}\right|^{2}$
was used to reduce the complexity of the expressions (note that by
definition $\sum_{i}x_{i}=\sum_{i}y_{i}=1$). It is straightforward
but tedious to carefully go through all cases in which the above vectors
are all aligned with each other, or null. Therefore a description
of the various possibilities is relegated to the appendix.

\begin{figure}
\begin{centering}
\includegraphics[scale=0.52]{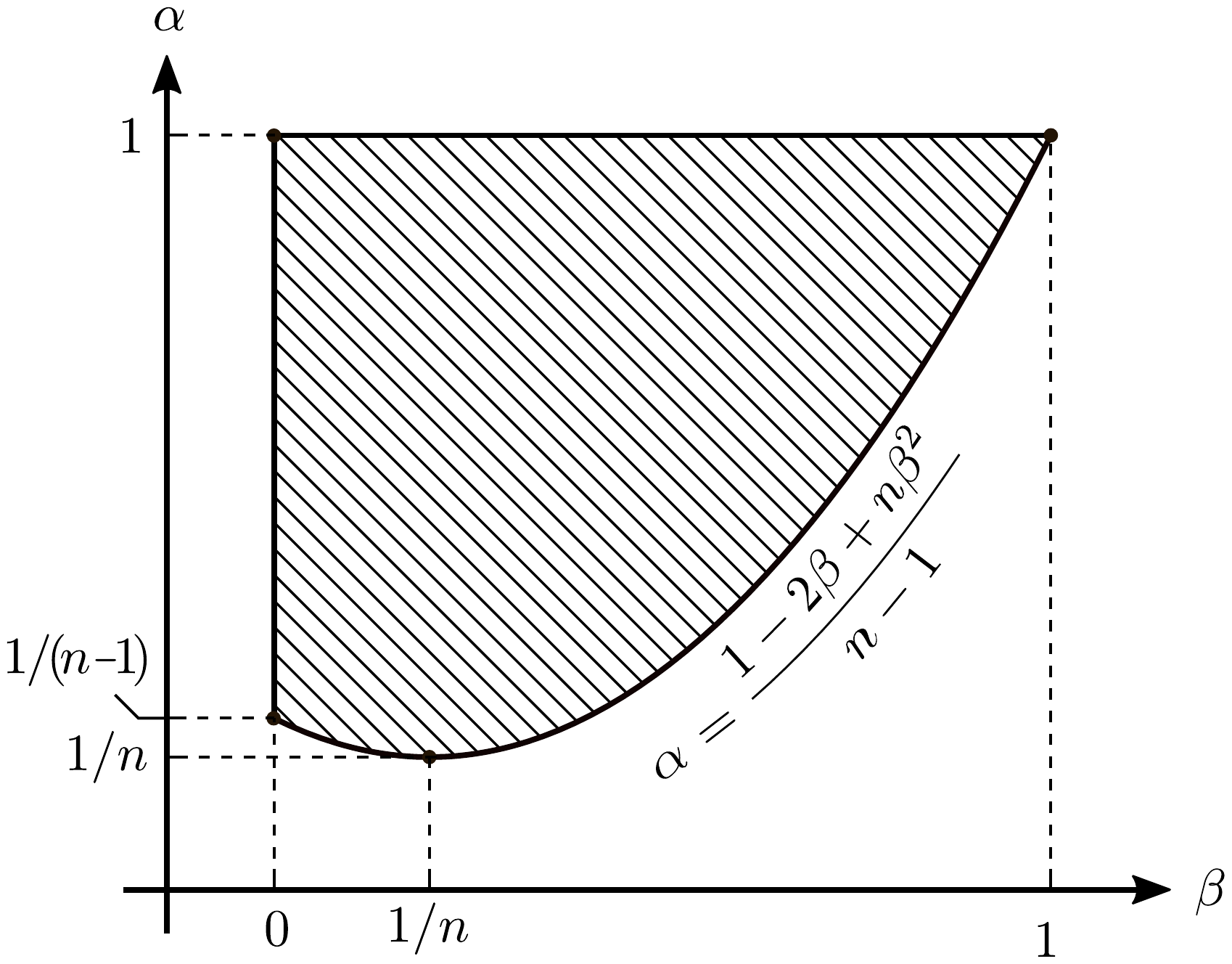}
\par\end{centering}
\caption{\label{fig:Allowed-ab}Allowed values of the important parameters
$\alpha$ and $\beta$, defined in equation (\ref{eq:alpha_beta}),
when $\Delta$ is symmetric.}
\end{figure}

The conclusion of the discussion contained therein is that the allowed
values of $\left(\alpha,\beta\right)$ correspond to the shaded area
in figure \ref{fig:Allowed-ab}, including the border lines. Note
that --- as expected --- this shape grows with $n$ since the $SU(n)$-invariant
potential can be seen as a special case of the $SU(n+1)$-invariant
where some field components are set to zero. As a consequence, the
BFB conditions on the $\lambda$'s become more stringent as $n$ increases.
Of particular relevance is the lower part of this shape, which is
defined by the quadratic relation
\begin{equation}
\alpha=\frac{1-2\beta+n\beta^{2}}{n-1}\,.\label{eq:lower-part-of-figure}
\end{equation}

The figure shown without proof in \cite{Bonilla:2015eha} corresponds
to the special situation where $n=2$, in which case the allowed region
for $\left(\alpha,\beta\right)$ is symmetric under reflection around
the vertical axis $\beta=1/2$; for $n>2$ there is a qualitative
change as the point $\left(\beta,\alpha\right)=\left(0,1/\left(n-1\right)\right)$
becomes distinct from $\left(0,1\right)$.

\section{\label{sec:4}The conditions for the $\lambda$'s}

We may now return to the inequalities in (\ref{eq:co-positivity-condition}).
Since they must hold for all $\alpha$ and $\beta$, substituting
$\alpha$ in $\lambda_{\Delta}+\alpha\lambda_{\Delta}^{\prime}>0$
by the smallest ($1/n$) and the largest (1) values this variable
can take, we conclude that this last inequality is equivalent to 
\begin{equation}
n\lambda_{\Delta}+\lambda_{\Delta}^{\prime}>0\textrm{ and }\lambda_{\Delta}+\lambda_{\Delta}^{\prime}>0\,.\label{eq:condtions-2}
\end{equation}
As observed already in \cite{Bonilla:2015eha}, the left-hand side
of $\lambda_{\phi\Delta}+\beta\lambda_{\phi\Delta}^{\prime}+\sqrt{\lambda_{\phi}\left(\lambda_{\Delta}+\alpha\lambda_{\Delta}^{\prime}\right)}>0$
is a monotonous function of both $\alpha$ and $\beta$, hence it
is enough that this condition holds on the border of the allowed $\alpha\beta$-region,
which is convex. In turn this is true if the inequality holds for
the points $\left(\beta,\alpha\right)=\left(0,1/\left(n-1\right)\right)$,
$\left(0,1\right)$, $\left(1,1\right)$ and the parabolic lower part
of the shaded region in figure (\ref{fig:Allowed-ab}). From the points
we get the constraints
\begin{equation}
\lambda_{\phi\Delta}+\sqrt{\lambda_{\phi}\left(\lambda_{\Delta}+\frac{\lambda_{\Delta}^{\prime}}{n-1}\right)}>0\textrm{ and }\lambda_{\phi\Delta}+\sqrt{\lambda_{\phi}\left(\lambda_{\Delta}+\lambda_{\Delta}^{\prime}\right)}>0\textrm{ and }\lambda_{\phi\Delta}+\lambda_{\phi\Delta}^{\prime}+\sqrt{\lambda_{\phi}\left(\lambda_{\Delta}+\lambda_{\Delta}^{\prime}\right)}>0\,.
\end{equation}

Five inequalities have so far been derived for the $\lambda$'s. The
second condition in expression (\ref{eq:co-positivity-condition})
must also hold for the parabolic lower part of the border, and that
constitutes the last problem to be dwelt with. In practice, we must
find the constraints on the quartic scalar couplings which make
\begin{equation}
f\left(\beta\right)\equiv\lambda_{\phi\Delta}+\beta\lambda_{\phi\Delta}^{\prime}+\sqrt{\lambda_{\phi}\left(\lambda_{\Delta}+\frac{1-2\beta+n\beta^{2}}{n-1}\lambda_{\Delta}^{\prime}\right)}
\end{equation}
positive for all $\beta\in\left[0,1\right]$. The sign of the second
derivative of this function does not change and in fact it is the
same as the one of $\lambda_{\Delta}^{\prime}$,
\begin{equation}
\textrm{sign}\left[f^{\prime\prime}\left(\beta\right)\right]=\textrm{sign}\left(\lambda_{\Delta}^{\prime}\right)\,,
\end{equation}
so $f$ has a single stationary point (where $f^{\prime}\left(\beta\right)=0$)
and it is an absolute minimum if $\lambda_{\Delta}^{\prime}>0$. Note
that if $\lambda_{\Delta}^{\prime}\leq0$ the value of $\lambda_{\phi\Delta}+\beta\lambda_{\phi\Delta}^{\prime}+\sqrt{\lambda_{\phi}\left(\lambda_{\Delta}+\alpha\lambda_{\Delta}^{\prime}\right)}$
is minimized instead for $\left(\beta,\alpha\right)=$$\left(0,1\right)$
or $\left(1,1\right)$, and both of these cases were already taken
into account above.

The final condition is then
\begin{equation}
f^{\prime}\left(0\right)>0\textrm{ or }f^{\prime}\left(1\right)<0\textrm{ or }\min\left[f\left(\beta\right)\right]>0\,,\label{eq:conditions-f-1}
\end{equation}
where $\min\left[f\left(\beta\right)\right]$ can be found by requiring
that $f^{\prime}\left(\beta\right)=0$ without caring if the value
of $\beta$ is between 0 and 1. In fact, the first two inequalities
in the expression above are necessary because if $f^{\prime}\left(0\right)$
is positive or $f^{\prime}\left(1\right)$ is negative the derivative
of $f\left(\beta\right)$ is null outside the interval $\beta\in\left[0,1\right]$.\footnote{In that case, the minimum of $f\left(\beta\right)$ in the $\left[0,1\right]$
interval is at one of the end-points ($\beta=0$ or 1). This corresponds
to the points $\left(\beta,\alpha\right)=\left(0,1/\left(n-1\right)\right)$
and $\left(1,1\right)$, which were already considered previously.} It is then rather simple to resolve the logical condition (\ref{eq:conditions-f-1})
is terms of $\lambda$'s.

In summary, the necessary and sufficient BFB condition for the $SU(n)$
invariant potential (\ref{eq:scalar-potential}) which have been derived
over the previous paragraphs is the following:
\begin{gather}
\lambda_{\phi}>0\;\textrm{ and }n\lambda_{\Delta}+\lambda_{\Delta}^{\prime}>0\textrm{ and }\lambda_{\Delta}+\lambda_{\Delta}^{\prime}>0\textrm{ and }\nonumber \\
\lambda_{\phi\Delta}+\sqrt{\lambda_{\phi}\left(\lambda_{\Delta}+\frac{\lambda_{\Delta}^{\prime}}{n-1}\right)}>0\textrm{ and }\lambda_{\phi\Delta}+\sqrt{\lambda_{\phi}\left(\lambda_{\Delta}+\lambda_{\Delta}^{\prime}\right)}>0\textrm{ and }\nonumber \\
\lambda_{\phi\Delta}+\lambda_{\phi\Delta}^{\prime}+\sqrt{\lambda_{\phi}\left(\lambda_{\Delta}+\lambda_{\Delta}^{\prime}\right)}>0\textrm{ and }\left[\lambda_{\phi\Delta}^{\prime}-\frac{1}{n-1}\frac{\lambda_{\Delta}^{\prime}\sqrt{\lambda_{\phi}}}{\sqrt{\lambda_{\Delta}+\frac{\lambda_{\Delta}^{\prime}}{n-1}}}>0\textrm{ or }\right.\nonumber \\
\left.\lambda_{\phi\Delta}^{\prime}+\frac{\lambda_{\Delta}^{\prime}\sqrt{\lambda_{\phi}}}{\sqrt{\lambda_{\Delta}+\lambda_{\Delta}^{\prime}}}<0\textrm{ or }n\lambda_{\phi\Delta}+\lambda_{\phi\Delta}^{\prime}+\sqrt{\left(n\frac{\lambda_{\Delta}}{\lambda_{\Delta}^{\prime}}+1\right)\left[n\lambda_{\Delta}^{\prime}\lambda_{\phi}-\left(n-1\right)\lambda_{\phi\Delta}^{\prime2}\right]}>0\right]\label{eq:BFB-symmetric-rep}
\end{gather}
This set of inequalities generalizes to any $SU(n)$ the somewhat
more compact formulae given in \cite{Bonilla:2015eha} for $n=2$.
The expression inside the square brackets corresponds to condition
(\ref{eq:conditions-f-1}); the first two square roots appearing in
it must be positive due to the other constraints (in particular (\ref{eq:condtions-2})).
On the other hand, if the first two conditions in the above OR expression
are false, then the argument of the last square root will always be
positive hence the full expression always makes sense.

\section{\label{sec:5}Other scalars}

\subsection{The 2-index anti-symmetric representation}

Let us now consider what happens if $\Delta$ transforms as the 2-index
anti-symmetric representation. The gauge transformation is the same
as in equation (\ref{eq:Delta-gauge-transformation}), hence the relevant
potential is the one given in expression (\ref{eq:scalar-potential}),
but now $\Delta$ is to be viewed as a generic $n\times n$ anti-symmetric
matrix. This feature makes it impossible to diagonize $\Delta$ with
a gauge transformation. One can however block-diagonalize it into
the form
\begin{equation}
\Delta=\textrm{diag}\left[\left(\begin{array}{cc}
0 & \Delta_{1}\\
-\Delta_{1} & 0
\end{array}\right),\left(\begin{array}{cc}
0 & \Delta_{2}\\
-\Delta_{2} & 0
\end{array}\right),\cdots,\left(\begin{array}{cc}
0 & \Delta_{\left\lfloor n/2\right\rfloor }\\
-\Delta_{\left\lfloor n/2\right\rfloor } & 0
\end{array}\right),\left(0\right)_{\textrm{if }n=\textrm{odd}}\right]
\end{equation}
where $\left\lfloor n/2\right\rfloor $ stands for the greatest integer
lesser than or equal to $n/2$. If $n$ is odd, there must be an extra
diagonal entry equal to 0. Nevertheless, the potential (\ref{eq:scalar-potential})
is only sensitive to the matrix combination $\Delta^{*}\Delta$ which
can be diagonalized:
\begin{equation}
\Delta^{*}\Delta=-\textrm{diag}\left(\left|\Delta_{1}\right|^{2},\left|\Delta_{1}\right|^{2},\left|\Delta_{2}\right|^{2},\left|\Delta_{2}\right|^{2},\cdots,\left|\Delta_{\left\lfloor n/2\right\rfloor }\right|^{2},\left|\Delta_{\left\lfloor n/2\right\rfloor }\right|^{2},0_{\textrm{if }n=\textrm{odd}}\right)\,.
\end{equation}
Two differences with the symmetric $\Delta$ can promptly be discerned:
\begin{enumerate}
\item There is an overall minus sign in $\Delta^{*}\Delta$. This can be
taken into account by swapping $\lambda_{\phi\Delta}$ and $\lambda_{\phi\Delta}^{\prime}$
by $-\lambda_{\phi\Delta}$ and $-\lambda_{\phi\Delta}^{\prime}$
in the BFB conditions. I will tacitly assume that this change has
been done from now on.
\item The eigenvalues of $\Delta^{*}\Delta$ appear repeated, except a zero
when $n$ is odd.
\end{enumerate}
Let us then consider first the case when $n$ is even. Using the notation
$n\equiv2n^{\prime}$ and $\left|\Phi_{i}\right|^{2}\equiv\left|\phi_{2i-1}\right|^{2}+\left|\phi_{2i}\right|^{2}$
we may write 
\begin{align}
\alpha & =\frac{\sum_{i}^{n^{\prime}}2\left|\Delta_{i}\right|^{4}}{\left(\sum_{i}^{n^{\prime}}2\left|\Delta_{i}\right|^{2}\right)^{2}}=\frac{1}{2}\frac{\sum_{i}^{n^{\prime}}\left|\Delta_{i}\right|^{4}}{\left(\sum_{i}^{n^{\prime}}\left|\Delta_{i}\right|^{2}\right)^{2}}\,,\label{eq:alpha-1/2}\\
\beta & =\frac{\sum_{i}^{n^{\prime}}\left|\Delta_{i}\right|^{2}\left(\left|\phi_{2i-1}\right|^{2}+\left|\phi_{2i}\right|^{2}\right)}{\left(\sum_{i}^{n^{\prime}}2\left|\Delta_{i}\right|^{2}\right)\left[\sum_{i}^{n^{\prime}}\left(\left|\phi_{2i-1}\right|^{2}+\left|\phi_{2i}\right|^{2}\right)\right]}\equiv\frac{1}{2}\frac{\sum_{i}^{n^{\prime}}\left|\Delta_{i}\right|^{2}\left|\Phi_{i}\right|^{2}}{\left(\sum_{i}^{n^{\prime}}\left|\Delta_{i}\right|^{2}\right)\left(\sum_{i}^{n^{\prime}}\left|\Phi_{i}\right|^{2}\right)}\,.\label{eq:beta-1/2}
\end{align}
Apart from the $1/2$ factors, these expressions are exactly what
one would have if $\Delta$ was a symmetric matrix with dimension
$n^{\prime}$. Hence, the allowed $\alpha\beta$-region is as depicted
in figure \ref{fig:Allowed-ab}, but shrunk by a factor of two in
both axis, and using $n^{\prime}=n/2$ instead of $n$. That means
that for $SU(n)$ the border of the figure goes through the points
$\left(0,1/\left(n-2\right)\right)$, $\left(1/n,1/n\right)$, $\left(0,1/2\right)$
and $\left(1/2,1/2\right)$. Based on these comments, it is rather
straightforward to make the necessary changes to the conditions (\ref{eq:BFB-symmetric-rep})
in order to obtain the BFB conditions when $\Delta$ is anti-symmetric
and $n$ is even (these are given explicitly below).

\begin{figure}
\begin{centering}
\includegraphics[scale=0.52]{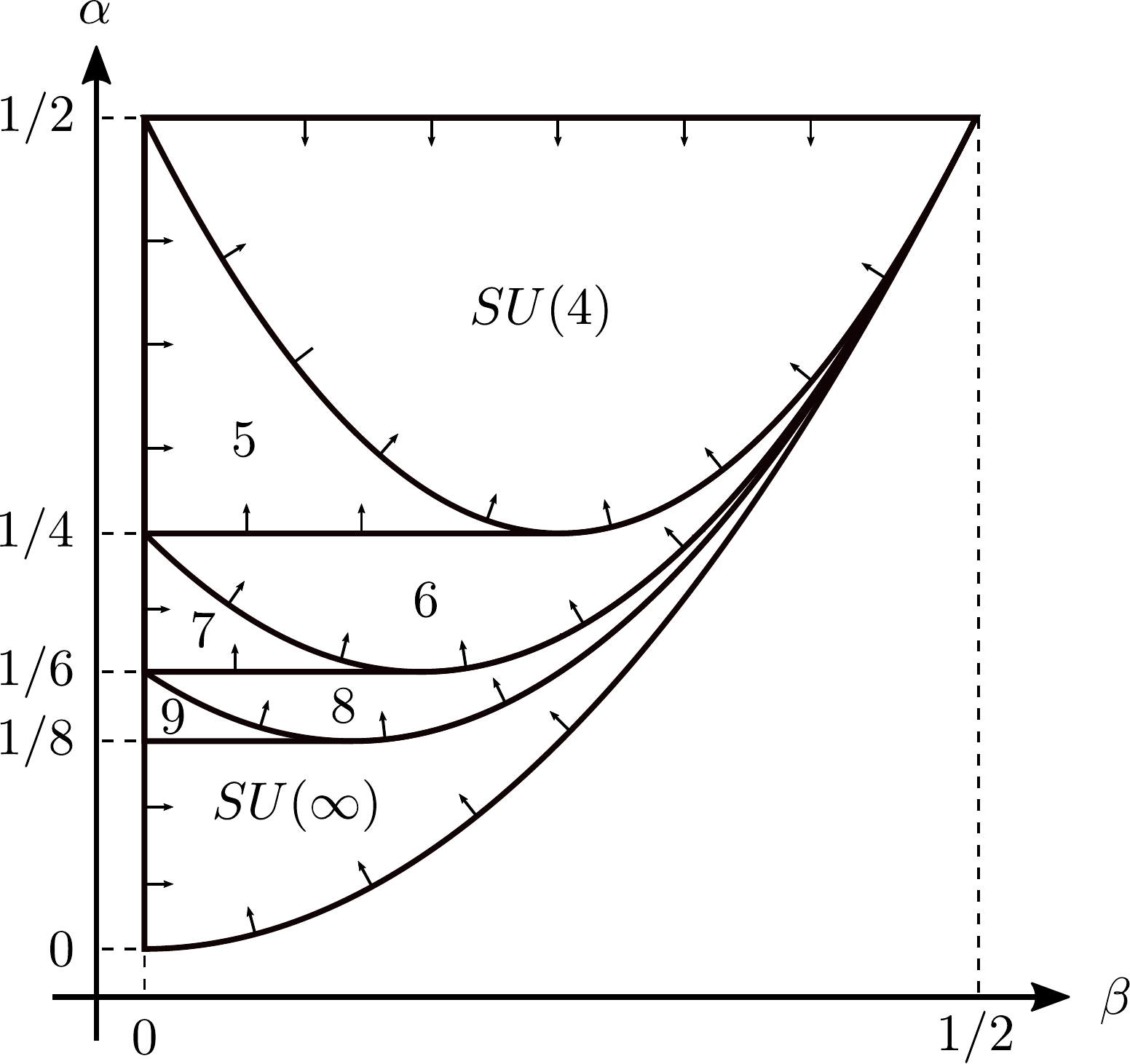}
\par\end{centering}
\caption{\label{fig:Allowed-ab-AntiS}The allowed region for the parameters
$\alpha$ and $\beta$, as defined in equations (\ref{eq:alpha-1/2}),
(\ref{eq:beta-1/2}) and (\ref{eq:beta-1/2-odd}), when $\Delta$
is anti-symmetric. The numbers shown refer to the $SU(n)$ group under
consideration. The shape of the allowed region is markedly different
for odd $n$'s when compared to even $n$'s; nevertheless the area
always increases with $n$.}
\end{figure}

When $n$ is odd, $\Delta^{*}\Delta$ contains an unpaired null eigenvalue,
which is an important feature. If we were to define $n\equiv2n^{\prime}+1$,
then $\alpha$ is as given in equation (\ref{eq:alpha-1/2}). However,
the denominator of $\beta$ now depends on $\left|\phi_{n}\right|$
while the numerator does not:
\begin{equation}
\beta=\frac{1}{2}\frac{\sum_{i}^{n^{\prime}}\left|\Delta_{i}\right|^{2}\left|\Phi_{i}\right|^{2}}{\left(\sum_{i}^{n^{\prime}}\left|\Delta_{i}\right|^{2}\right)\left(\sum_{i}^{n^{\prime}}\left|\Phi_{i}\right|^{2}+\left|\phi_{n}\right|^{2}\right)}\,.\label{eq:beta-1/2-odd}
\end{equation}
This is a decreasing function of $\left|\phi_{n}\right|$, reaching
a maximum given by equation (\ref{eq:beta-1/2}) (when $\left|\phi_{n}\right|=0$)
and a minimum of $0$ when $\left|\phi_{n}\right|\rightarrow\infty$.
Therefore, compared to figure \ref{fig:Allowed-ab}, the allowed $\alpha\beta$-region
shrinks by a factor of two in both axis and $n^{\prime}$ replaces
$n$. Furthermore, for all values of $\alpha$ ($1/\left(2n^{\prime}\right)$
to $1/2$) $\beta$ can be null, which means that in $\left(\beta,\alpha\right)$
coordinates, a straight line connecting $\left(0,1/2n^{\prime}\right)$
to $\left(1/2n^{\prime},1/2n^{\prime}\right)$ forms part of the border
of the allowed space. Figure \ref{fig:Allowed-ab-AntiS} shows some
examples.

Note that the cases $n=2,3$ are exceptional, since $n^{\prime}$
is 1 and $\alpha$ has a fixed value of $1/2$. In other words $\textrm{Tr}\left(\Delta^{*}\Delta\Delta^{*}\Delta\right)=\left[\textrm{Tr}\left(\Delta^{*}\Delta\right)\right]^{2}/2$
and therefore $V^{(4)}$ contains only 4 independent coupling (it
depends on $\lambda_{\Delta}$ and $\lambda_{\Delta}^{\prime}$ only
through the combination $\lambda_{\Delta}+\lambda_{\Delta}^{\prime}/2$).
For $n=2$, $\beta$ also has the fixed value $1/2$, while for $n=3$
it can be any number between 0 and $1/2$.

Taking into account the above considerations, the BFB condition in
(\ref{eq:BFB-symmetric-rep}) for the symmetric representation is
modified to the following form, which is valid for all values of $n$,
regardless of its parity. First define $\widetilde{n}$ to be the
largest even integer smaller or equal to $n$: $\widetilde{n}=n$
if $n$ is even, otherwise $\widetilde{n}=n-1$. Then for $n>3$ the
BFB conditions are the following:
\begin{gather}
\lambda_{\phi}>0\;\textrm{ and }\widetilde{n}\lambda_{\Delta}+\lambda_{\Delta}^{\prime}>0\textrm{ and }2\lambda_{\Delta}+\lambda_{\Delta}^{\prime}>0\textrm{ and }\nonumber \\
-\lambda_{\phi\Delta}+\sqrt{\lambda_{\phi}\left(\lambda_{\Delta}+\frac{\lambda_{\Delta}^{\prime}}{2n-\widetilde{n}-2}\right)}>0\textrm{ and }-\lambda_{\phi\Delta}+\sqrt{\lambda_{\phi}\left(\lambda_{\Delta}+\frac{\lambda_{\Delta}^{\prime}}{2}\right)}>0\textrm{ and }\nonumber \\
-\lambda_{\phi\Delta}-\frac{\lambda_{\phi\Delta}^{\prime}}{2}+\sqrt{\lambda_{\phi}\left(\lambda_{\Delta}+\frac{\lambda_{\Delta}^{\prime}}{2}\right)}>0\textrm{ and }\left[-\lambda_{\phi\Delta}^{\prime}-\frac{1}{\widetilde{n}-2}\frac{2\lambda_{\Delta}^{\prime}\sqrt{\lambda_{\phi}}}{\sqrt{\lambda_{\Delta}+\frac{\lambda_{\Delta}^{\prime}}{\widetilde{n}-2}}}>0\textrm{ or }\right.\nonumber \\
\left.-\lambda_{\phi\Delta}^{\prime}+\frac{\lambda_{\Delta}^{\prime}\sqrt{\lambda_{\phi}}}{\sqrt{\lambda_{\Delta}+\frac{\lambda_{\Delta}^{\prime}}{2}}}<0\textrm{ or }-\widetilde{n}\lambda_{\phi\Delta}-\lambda_{\phi\Delta}^{\prime}+\sqrt{\left(\widetilde{n}\frac{\lambda_{\Delta}}{\lambda_{\Delta}^{\prime}}+1\right)\left[\widetilde{n}\lambda_{\Delta}^{\prime}\lambda_{\phi}-\left(\frac{\widetilde{n}}{2}-1\right)\lambda_{\phi\Delta}^{\prime2}\right]}>0\right]\label{eq:BFB-anti-symmetric-rep}
\end{gather}
For $n=2$ ($\Delta$ is an $SU(2)$ singlet) the conditions are
\begin{gather}
\lambda_{\phi}>0\textrm{ and }2\lambda_{\Delta}+\lambda_{\Delta}^{\prime}>0\textrm{ and }-\lambda_{\phi\Delta}-\frac{\lambda_{\phi\Delta}^{\prime}}{2}+\sqrt{\lambda_{\phi}\left(\lambda_{\Delta}+\frac{\lambda_{\Delta}^{\prime}}{2}\right)}>0\,,
\end{gather}
while for $n=3$ ($\Delta^{*}$ is an $SU(3)$ triplet) it is additionally
necessary that
\begin{equation}
-\lambda_{\phi\Delta}+\sqrt{\lambda_{\phi}\left(\lambda_{\Delta}+\frac{\lambda_{\Delta}^{\prime}}{2}\right)}>0\,.
\end{equation}

\subsection{The adjoint representation}

We may move on to the significantly more elaborate case where $\Delta$
transforms as an adjoint representation $\Delta_{\,\,j}^{i}$:
\begin{gather}
\Delta\rightarrow U\Delta U^{\dagger}\,.\label{eq:G-gauge-transformation}
\end{gather}
This $\Delta$ can be viewed as a traceless hermitian matrix with
$n^{2}-1$ real degrees of freedom. Reusing the same names for the
$\lambda$ quartic couplings, the most general $SU(n)$ invariant
potential can be written as
\begin{align}
V^{(4)} & =\frac{\lambda_{\phi}}{2}\left(\phi^{\dagger}\phi\right)^{2}+\frac{\lambda_{\Delta}}{2}\left[\textrm{Tr}\left(\Delta^{2}\right)\right]^{2}+\frac{\lambda_{\Delta}^{\prime}}{2}\textrm{Tr}\left(\Delta^{4}\right)+\lambda_{\phi\Delta}\left(\phi^{\dagger}\phi\right)\textrm{Tr}\left(\Delta^{2}\right)+\lambda_{\phi\Delta}^{\prime}\phi^{\dagger}\Delta\Delta\phi\label{eq:scalar-potential-G}
\end{align}
which is an expression somewhat similar to the one in equation (\ref{eq:scalar-potential}).
With a gauge transformation it is always possible to diagonalize $\Delta$,
however unlike when $\Delta$ was symmetric, the matrix must remain
traceless:\footnote{The reader might be puzzled by the fact that in the case of $SU(2)$,
the adjoint and the 2-index symmetric representations are the same.
Yet the text implies that if we treat $\Delta$ as a symmetric matrix
(let us call it $\Delta_{S}$), the best that can be done with the
gauge symmetry is to cast it in a diagonal form (two real degrees
of freedom), while $\Delta$ seen as a traceless hermitian matrix
($\Delta_{H}$) can be reduced to a real traceless diagonal matrix,
with only one real degree of freedom. The reason behind this apparent
contradiction is that $\Delta_{S}$ may represent a complex triplet,
while $\Delta_{H}$ must stand for a real triplet, with half of the
degrees of freedom to start with. Even if we take $\Delta_{S}$ to
be a real matrix, the two cases would still be inequivalent due to
a different choice of basis (as can be seen from the fact that $\Delta_{S}\epsilon$
is not hermitian, with $\epsilon$ being the Levi-Civita matrix).}
\begin{equation}
\Delta=\textrm{diag}\left(\Delta_{1},\Delta_{2},\cdots,\Delta_{n-1},-\Delta_{1}-\Delta_{2}\cdots-\Delta_{n-1}\right)\,.
\end{equation}
This leads to non-trivial complications in the analysis of $V^{(4)}$,
as was pointed out in \cite{Frautschi:1981jh}. We may define $\alpha$
and $\beta$ as before (see equation (\ref{eq:alpha_beta})), with
the understanding that $\Delta_{n}=-\Delta_{1}-\Delta_{2}\cdots-\Delta_{n-1}$,
and try to find the allowed values of these two variables. The authors
of \cite{Frautschi:1981jh} conjectured that the configurations associated
to the border of the valid $\alpha\beta$-space are those of the form
\begin{align}
\phi & =\left(0,0,\cdots,0,1\right)^{T}\,,\\
\Delta & =\textrm{diag}\left(\underbrace{a,\cdots,a}_{m_{1}},\underbrace{b,\cdots,b}_{m_{2}},-am_{1}-bm_{2}\right)\,
\end{align}
plus some lesser important cases to be discussed later.\footnote{Numerical scans suggest that this conjecture is true.}
Note that $n=m_{1}+m_{2}+1$, so for a fixed $n$ only one of the
integers $m_{1,2}$ can be picked freely (for definiteness I'll take
$m_{1}$ as the independent variable). We get the following relation
between $\alpha$ and $\beta$ for this particular VEV configuration,
with $a$ and $b$ eliminated:
\begin{align}
\alpha & =\beta^{2}\left(1+A+2B+C\right)-2\beta\left(A+B\right)+A\nonumber \\
 & \pm\frac{4\left(m_{1}-m_{2}\right)}{\left(m_{1}+m_{2}\right)^{3}}\sqrt{\frac{\beta}{m_{1}m_{2}}}\left[m_{1}+m_{2}-\left(1+m_{1}+m_{2}\right)\beta\right]^{3/2}\label{eq:adjoint-curves}
\end{align}
with
\begin{equation}
A\equiv\frac{m_{1}^{2}-m_{1}m_{2}+m_{2}^{2}}{m_{1}m_{2}\left(m_{1}+m_{2}\right)},\;B\equiv\frac{m_{1}^{2}-4m_{1}m_{2}+m_{2}^{2}}{m_{1}m_{2}\left(m_{1}+m_{2}\right)^{2}},\;C\equiv\frac{m_{1}^{2}-6m_{1}m_{2}+m_{2}^{2}}{m_{1}m_{2}\left(m_{1}+m_{2}\right)^{3}}\,.
\end{equation}
There are two choices for each choice of $m_{1}$, depending on the
sign selected for the last term in the $\alpha$ expression, but it
is sufficient to always pick the plus sign, as the minus sign can
be replicated by swapping $m_{1}$ and $m_{2}$ ($m_{1}\rightarrow n-1-m_{1}$).
Unlike when $\Delta$ was symmetric (or skew-symmetric), the border
of the $\alpha\beta$-space is no longer composed exclusively of straight
lines and a parabola; now the relation between $\alpha$ and $\beta$
is significantly more complicated and furthermore one should consider
more than a single curve, since $m_{1}$ can take values from 1 to
$n-2$. One might have hoped that a single $m_{1}$ is relevant for
the demarcation of the border line, but this is not the case: several
of them contribute, each for some specific range of $\beta$.

\begin{figure}
\begin{centering}
\includegraphics[scale=0.5]{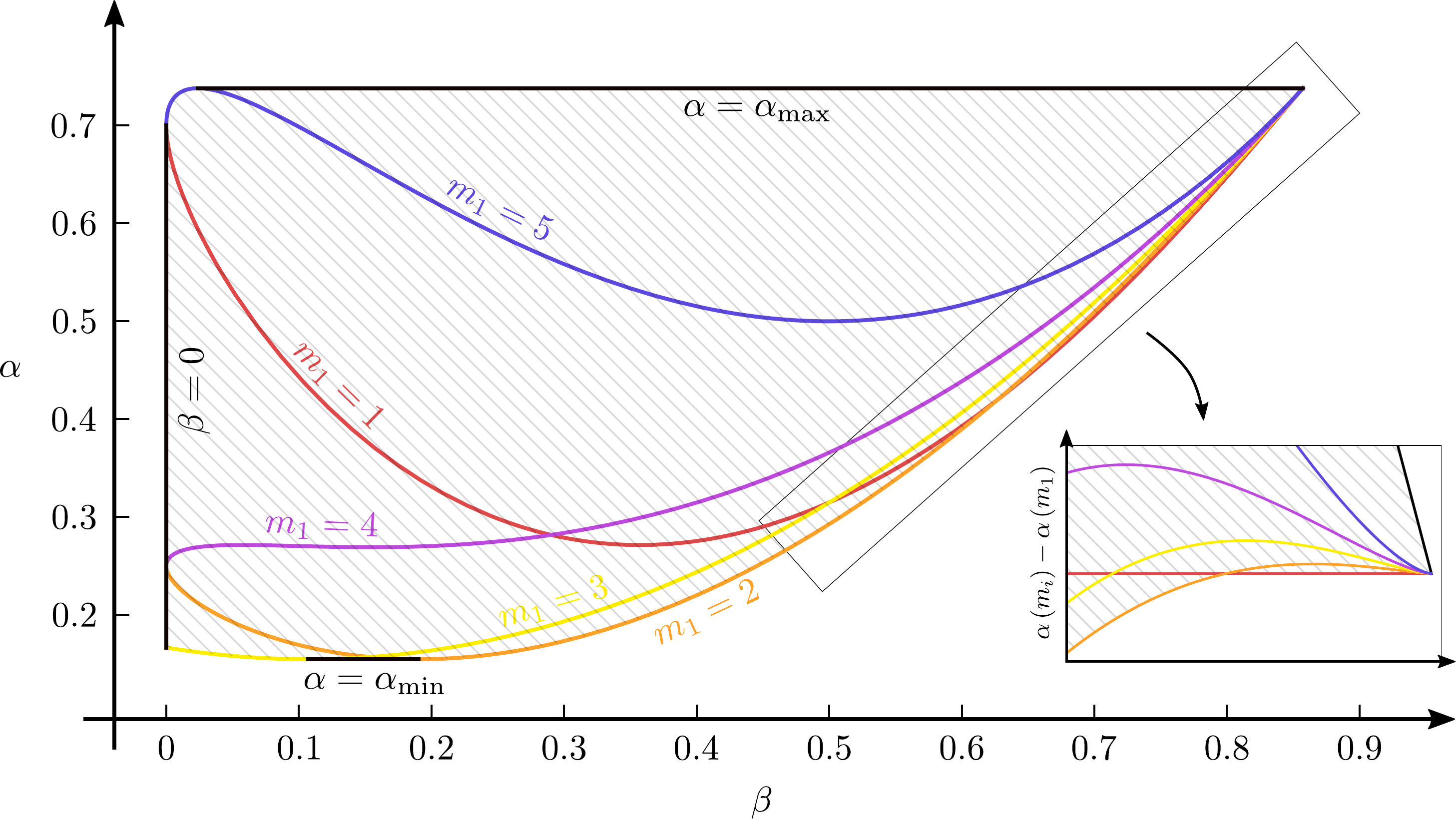}
\par\end{centering}
\caption{\label{fig:Allowed-ab-Adjoint}Demarcation lines of the allowed $\alpha\beta$-space
(shaded area) for an $SU(7)$ invariant potential with a fundamental
and an adjoint representation. The curved lines (in color) follow
equation (\ref{eq:adjoint-curves}), while the straight ones (in black)
are described by the expressions (\ref{eq:adjoint-straight-lines-1})--(\ref{eq:adjoint-straight-lines-2}),
obtained in reference \cite{Frautschi:1981jh}. The inlet clarifies
what is happening on the right side of the plot, with the $m_{1}=1,2,3$
curves all being important for the demarcation of the bottom border
line.}
\end{figure}

Figure \ref{fig:Allowed-ab-Adjoint} illustrates what happens for
$SU(7)$ (that is $n=7$). One can see there that the border line
is also made-up of horizontal and vertical straight lines (see \cite{Frautschi:1981jh});
nevertheless they are irrelevant for the stability of the vacuum.\footnote{The reason is as follows. We need to find the minimum of the expressions
appearing in the inequalities (\ref{eq:co-positivity-condition})
however, since these expressions are monotonous functions of $\alpha$
and $\beta$, one can disregard straight portions of the $\alpha\beta$-border
line (it is enough to consider their endpoints where the expressions
will always reach a minimum).} Noting that $\beta\in\left[0,\left(n-1\right)/n\right]$ and $\alpha\in\left[\alpha_{\textrm{min}},\alpha_{\textrm{max}}\right]$
with
\begin{gather}
\alpha_{\textrm{min}}=\begin{cases}
\frac{1}{n} & n\textrm{ even}\\
\frac{n^{2}+3}{n\left(n^{2}-1\right)} & n\textrm{ odd}
\end{cases}\,,\label{eq:alpha-min}\\
\alpha_{\textrm{max}}=\frac{\left(n-1\right)^{3}+1}{\left(n-1\right)n^{2}}\,,\label{eq:alpha-max}
\end{gather}
there are the following straight lines:
\begin{align}
\alpha=\alpha_{\textrm{min}} & \textrm{ and }\beta\in\left[\frac{n-1}{n\left(n+1\right)},\frac{n+1}{n\left(n-1\right)}\right]\textrm{ (line exists only for even }n\textrm{)}\,,\label{eq:adjoint-straight-lines-1}\\
\alpha=\alpha_{\textrm{max}} & \textrm{ and }\beta\in\left[\frac{1}{n\left(n-1\right)},\frac{n-1}{n}\right]\,,\\
\beta=0 & \textrm{ and }\alpha\in\begin{cases}
\left[\frac{n^{2}-2n+4}{n^{3}-3n^{2}+2n},\frac{n^{2}-5n+7}{\left(n-2\right)\left(n-1\right)}\right] & n\textrm{ even}\\
\left[\frac{1}{n-1},\frac{n^{2}-5n+7}{\left(n-2\right)\left(n-1\right)}\right] & n\textrm{ odd}
\end{cases}\,.\label{eq:adjoint-straight-lines-2}
\end{align}

Since the shape of the $\alpha\beta$-space is quite elaborate, we
may focus instead on the rectangle containing it and derive the following
simple but potentially useful BFB condition --- which is sufficient
but not necessary for vacuum stability. It consists on demanding that
all the following expressions are positive:
\begin{gather}
\lambda_{\phi},\;\lambda_{\Delta}+\alpha_{\textrm{min}/\textrm{max}}\lambda_{\Delta}^{\prime},\;\lambda_{\phi\Delta}+\beta_{\textrm{min}/\textrm{max}}\lambda_{\phi\Delta}^{\prime}+\sqrt{\lambda_{\phi}\left(\lambda_{\Delta}+\alpha_{\textrm{min}/\textrm{max}}\lambda_{\Delta}^{\prime}\right)}\,.
\end{gather}
One should take every combination of $\alpha$ and $\beta$ at their
minimum and maximum values (see equations (\ref{eq:alpha-min}), (\ref{eq:alpha-max})
and the text immediately preceding them), hence there is a total of
$1+2+4=7$ quantities to be checked.

\section{\label{sec:6}The 1-2-3 $SU(2)$ potential}

Neutrino masses can be generated at tree level by introducing in the
Standard Model a scalar $\Delta$ with the $SU(2)_{L}\times U(1)_{Y}$
quantum numbers $\left(\boldsymbol{3},1\right)$. Via the seesaw type-II
mechanism, neutrinos acquire a mass $m_{\nu}=Y_{\nu}\mu\left\langle \phi\right\rangle ^{2}/m_{\Delta^{0}}^{2}$
where
\begin{itemize}
\item $Y_{\nu}$ is the Yukawa coupling matrix regulating the interaction
$L_{i}^{T}CL_{j}\Delta$ between left-handed leptons and $\Delta$;
\item $m_{\Delta^{0}}$ stands for the mass of the neutral component of
$\Delta$;
\item $\mu$ is a mass which controls the strength of the trillinear interaction
$\phi^{\dagger}\Delta\phi^{*}$ between $\Delta$ and the Higgs doublet
$\phi$.
\end{itemize}
Note that lepton number is restored in the limit where $\mu$ vanishes,
so this symmetry protects $\mu$ from big radiative corrections, and
that is why the smallness of $m_{\nu}$ is usually attributed to the
tiny value of this mass parameter.

\begin{figure}
\begin{centering}
\includegraphics[scale=0.9]{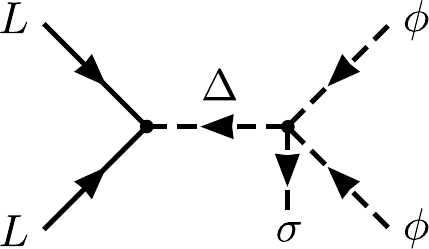}
\par\end{centering}
\caption{\label{fig:Neutrino-mass}Neutrino mass diagram in the 1-2-3 model.
When $\sigma$ acquires a non-zero vacuum expectation value, the $LL\phi\phi$
Weinberg operator \cite{Weinberg:1979sa} is generated ($L$ and $\phi$
represent the left-handed leptons and the Higgs doublet).}
\end{figure}

As an alternative, lepton number might be spontaneously violated.
To that end one can introduce a scalar singlet with no hypercharge
and two units of lepton number \cite{Schechter:1981cv}, so that an
interaction $\frac{\lambda_{\sigma\phi\Delta}}{2}\sigma\phi^{\dagger}\Delta\phi^{*}+\textrm{h.c.}$
is allowed by all symmetries; once this scalar acquires a vacuum
expectation value, an effective $\mu$ equal to $\lambda_{\sigma\phi\Delta}\left\langle \sigma\right\rangle $
is generated (see figure \ref{fig:Neutrino-mass}). 
With a singlet $\sigma$ ($\boldsymbol{1}$), a doublet $\phi$ ($\boldsymbol{2}$)
and a triplet $\Delta$ ($\boldsymbol{3}$), this setup is sometimes
called the 1-2-3 model. The full scalar potential reads

\begin{align}
V^{(4)}\left(\phi,\Delta,\sigma\right) & =V^{(4)}\left(\phi,\Delta\right)+\frac{\lambda_{\sigma}}{2}\left|\sigma\right|^{4}+\lambda_{\sigma\phi}\left|\sigma\right|^{2}\phi^{\dagger}\phi+\lambda_{\sigma\Delta}\left|\sigma\right|^{2}\textrm{Tr}\left(\Delta\Delta^{*}\right)+\left(\frac{\lambda_{\sigma\phi\Delta}}{2}\sigma\phi^{\dagger}\Delta\phi^{*}+\textrm{h.c.}\right)\label{eq:1-2-3-potential}
\end{align}
where $V^{(4)}\left(\phi,\Delta\right)$ contains only terms with
$\phi$ and $\Delta$ and was given previously in equation (\ref{eq:scalar-potential}).
Once again a gauge transformation can be used to diagonalize $\Delta$
($\rightarrow\textrm{diag}\left(\Delta_{1},\Delta_{2}\right)$), in
which case we may make the replacements $\phi^{\dagger}\phi\rightarrow\left|\phi_{1}\right|^{2}+\left|\phi_{2}\right|^{2}$,
$\textrm{Tr}\left(\Delta\Delta^{*}\right)\rightarrow\left|\Delta_{1}\right|^{2}+\left|\Delta_{2}\right|^{2}$
and $\phi^{\dagger}\Delta\phi^{*}\rightarrow\left(\phi_{1}^{*}\right)^{2}\Delta_{1}+\left(\phi_{2}^{*}\right)^{2}\Delta_{2}$.
This last expression is the only one sensitive to the phases of the
fields, so the potential above is minimal when
\begin{equation}
\frac{\lambda_{\sigma\phi\Delta}}{2}\sigma\phi^{\dagger}\Delta\phi^{*}+\textrm{h.c.}=-\left|\lambda_{\sigma\phi\Delta}\right|\left|\sigma\right|\left(\left|\phi_{1}\right|^{2}\left|\Delta_{1}\right|+\left|\phi_{2}\right|^{2}\left|\Delta_{2}\right|\right)\,.\label{eq:sigma-delta-phi-term}
\end{equation}
We have seen that $V^{(4)}\left(\phi,\Delta\right)$ depends only
on 4 field components --- $\left|\phi_{1,2}\right|$ and $\left|\Delta_{1,2}\right|$
--- or equivalently $\left|\phi_{1}\right|^{2}+\left|\phi_{2}\right|^{2}$,
$\left|\Delta_{1}\right|^{2}+\left|\Delta_{2}\right|^{2}$, $\alpha$
and $\beta$ (see equation (\ref{eq:alpha_beta})). With the introduction
of $\sigma$, the minimum of the potential will depend only on one
extra field $\left|\sigma\right|$,\footnote{In analogy with $\alpha$ and $\beta$, we may define the variable
\[
\gamma\equiv\frac{\left|\phi_{1}\right|^{2}\left|\Delta_{1}\right|+\left|\phi_{2}\right|^{2}\left|\Delta_{2}\right|}{\left(\left|\phi_{1}\right|^{2}+\left|\phi_{2}\right|^{2}\right)\sqrt{\left|\Delta_{1}\right|^{2}+\left|\Delta_{2}\right|^{2}}}\,.
\]
Nevertheless, $\gamma$ can be written as a function of $\alpha$
and $\beta$ so it does not constitute an independent degree of freedom.} nevertheless the potential itself becomes significantly more complicated,
with 4 new $\lambda$'s. In fact, to find the BFB conditions of the
1-2-3 potential it would be necessary to minimize a polynomial with
a quadratic dependence on the $\left|\phi_{i}\right|^{2}$ and crucially
a quartic dependence on the variables $\left|\sigma\right|$ and $\left|\Delta_{i}\right|$.
The results on the copositivity of quadratic functions cannot be used
here, and one can appreciate from \cite{2019arXiv190900880S,Kannike:2016fmd}
that handling multi-variable quartic functions is very complicated,
hence it seems unwise to try to find the necessary and sufficient
BFB of the potential in equation (\ref{eq:1-2-3-potential}).\footnote{Neglecting the special case when $\sigma=0$ (which was already addressed),
one can make the variable substitution $\left|\Delta_{i}\right|\rightarrow\left|\sigma\right|\left|\Delta_{i}^{\prime}\right|$,
turning the potential into a quadratic function of $\left|\phi_{1}\right|^{2}$,
$\left|\phi_{2}\right|^{2}$ and $\left|\sigma\right|^{2}$, hence
the known copositivity results can be applied to these three variables.
The result is a complicated system of inequalities involving $\left|\Delta_{1}^{\prime}\right|$
and $\left|\Delta_{2}^{\prime}\right|$ which would still need to
be resolved for all values of these variables. Nevertheless, for a
numerical check of whether or not a specific potential is bounded
from below, those inequalities might be of some use since for each
set of $\lambda$'s one only has to sample a 2-dimensional field space
rather the original 12-dimensional one.} However, for the study of neutrino masses in the 1-2-3 model it might
be good enough to find some acceptable values of the $\lambda$'s
(not necessarily all of them). 

One important case is when the coupling $\lambda_{\sigma\phi\Delta}$
is too small to be relevant for the stability of the vacuum. The neutrino
mass matrix is given by the formula $Y_{\nu}\lambda_{\sigma\phi\Delta}^{*}\left\langle \sigma\right\rangle \left\langle \phi\right\rangle ^{2}/m_{\Delta^{0}}^{2}$
with $m_{\Delta^{0}}$ often taken to be quite low --- of the TeV
order --- so the product $Y_{\nu}\lambda_{\sigma\phi\Delta}^{*}\left\langle \sigma\right\rangle $
must be tiny. Therefore the approximation $\lambda_{\sigma\phi\Delta}\approx0$
is an important and well motivated one. Without this coupling, the
1-2-3 potential becomes a quadratic function of the non-negative variables
$\left|\phi_{1,2}\right|^{2}$, $\left|\Delta_{1,2}\right|^{2}$ and
$\left|\sigma\right|^{2}$, hence the potential is bounded from below
if and only if the symmetric matrix

\begin{equation}
\left(\begin{array}{ccccc}
\lambda_{\phi} & \lambda_{\phi} & \lambda_{\phi\Delta}+\lambda_{\phi\Delta}^{\prime} & \lambda_{\phi\Delta} & \lambda_{\sigma\phi}\\
\cdot & \lambda_{\phi} & \lambda_{\phi\Delta} & \lambda_{\phi\Delta}+\lambda_{\phi\Delta}^{\prime} & \lambda_{\sigma\phi}\\
\cdot & \text{\ensuremath{\cdot}} & \lambda_{\Delta}+\lambda_{\Delta}^{\prime} & \lambda_{\Delta} & \lambda_{\sigma\Delta}\\
\cdot & \cdot & \cdot & \lambda_{\Delta}+\lambda_{\Delta}^{\prime} & \lambda_{\sigma\Delta}\\
\text{\ensuremath{\cdot}} & \cdot & \cdot & \cdot & \lambda_{\sigma}
\end{array}\right)
\end{equation}
is co-positive. It is straightforward to obtain the explicit set of
inequalities which the $\lambda$'s must obey (for example with the
method described in \cite{COTTLE1970295}; see also \cite{Kannike:2012pe}),
however I will not reproduce the expressions here since they are long
and not very instructive.

If $\left|\lambda_{\sigma\phi\Delta}\right|$ is sizable one might
consider the following strategy. For any scalar field configuration,
it is either true that $\left|\sigma\right|\geq\sqrt{\left|\Delta_{1}\right|^{2}+\left|\Delta_{2}\right|^{2}}$
or the opposite, hence
\begin{equation}
-\left|\sigma\right|\left(\left|\phi_{1}\right|^{2}\left|\Delta_{1}\right|+\left|\phi_{2}\right|^{2}\left|\Delta_{2}\right|\right)\leq-\left|\sigma\right|^{2}\left(\left|\phi_{1}\right|^{2}+\left|\phi_{2}\right|^{2}\right)\textrm{ or }-\left(\left|\Delta_{1}\right|^{2}+\left|\Delta_{2}\right|^{2}\right)\left(\left|\phi_{1}\right|^{2}+\left|\phi_{2}\right|^{2}\right)
\end{equation}
By replacing in the potential $V^{(4)}\left(\phi,\Delta,\sigma\right)$
the left term with the terms on the right, we get two potentials,
both of which depend only on $\left|\phi_{1,2}\right|^{2}$, $\left|\Delta_{1,2}\right|^{2}$
and $\left|\sigma\right|^{2}$. Therefore the 1-2-3 potential is bounded
from below if both the following symmetric matrices are co-positive:
\begin{gather}
\left(\begin{array}{ccccc}
\lambda_{\phi} & \lambda_{\phi} & \lambda_{\phi\Delta}+\lambda_{\phi\Delta}^{\prime} & \lambda_{\phi\Delta} & \lambda_{\sigma\phi}-\left|\lambda_{\sigma\phi\Delta}\right|\\
\cdot & \lambda_{\phi} & \lambda_{\phi\Delta} & \lambda_{\phi\Delta}+\lambda_{\phi\Delta}^{\prime} & \lambda_{\sigma\phi}-\left|\lambda_{\sigma\phi\Delta}\right|\\
\cdot & \text{\ensuremath{\cdot}} & \lambda_{\Delta}+\lambda_{\Delta}^{\prime} & \lambda_{\Delta} & \lambda_{\sigma\Delta}\\
\cdot & \cdot & \cdot & \lambda_{\Delta}+\lambda_{\Delta}^{\prime} & \lambda_{\sigma\Delta}\\
\text{\ensuremath{\cdot}} & \cdot & \cdot & \cdot & \lambda_{\sigma}
\end{array}\right)\,,\\
\left(\begin{array}{ccccc}
\lambda_{\phi} & \lambda_{\phi} & \lambda_{\phi\Delta}+\lambda_{\phi\Delta}^{\prime}-\left|\lambda_{\sigma\phi\Delta}\right| & \lambda_{\phi\Delta}-\left|\lambda_{\sigma\phi\Delta}\right| & \lambda_{\sigma\phi}\\
\cdot & \lambda_{\phi} & \lambda_{\phi\Delta}-\left|\lambda_{\sigma\phi\Delta}\right| & \lambda_{\phi\Delta}+\lambda_{\phi\Delta}^{\prime}-\left|\lambda_{\sigma\phi\Delta}\right| & \lambda_{\sigma\phi}\\
\cdot & \text{\ensuremath{\cdot}} & \lambda_{\Delta}+\lambda_{\Delta}^{\prime} & \lambda_{\Delta} & \lambda_{\sigma\Delta}\\
\cdot & \cdot & \cdot & \lambda_{\Delta}+\lambda_{\Delta}^{\prime} & \lambda_{\sigma\Delta}\\
\text{\ensuremath{\cdot}} & \cdot & \cdot & \cdot & \lambda_{\sigma}
\end{array}\right)\,.
\end{gather}
Note however that this is not a necessary condition: the 1-2-3 potential
might be bounded from below even if it fails to pass this test.

\section{\label{sec:7}Conclusions}

Scalar potentials are quartic functions of several field components,
hence their analysis can be quite complicated. That is why it is only
possible to write down the necessary and sufficient conditions for
these functions to be bounded from below in simple cases, when the
number of scalar representations is small. In this work, I have derived
these constraints for $SU(n)$ invariants potentials with two fields:
one transforming under the fundamental representation and the other
as a 2-index representation (the symmetric or the anti-symmetric one).
The case where the 2-index representation is the adjoint is substantially
more complicated, hence I have only provided in a closed form a sufficient
condition for the potential to be stable.

The combination of fields above mentioned appears in several models
extending the Standard Model gauge group. The special case where $n=2$
and the scalars are a doublet and a triplet is particularly important
because these fields participate in the seesaw type-II mechanism which
might be responsible for neutrino mass generation. The BFB conditions
for this scenario were already presented in \cite{Bonilla:2015eha},
although a crucial step necessary to derive this result was not shown
explicitly, as the relevant calculations were performed with a computer
algebra system. In this work, I have provided a fully analytical proof
of this result, which is valid for any $SU(n)$ group.

One can also add a scalar singlet to the Standard Model on top of
the triplet used in the type-II seesaw mechanism. With the introduction
of the singlet, lepton number can be broken spontaneously rather than
explicitly, leading to important phenomenological consequences. Yet
the scalar potential of this so-called 1-2-3 model contains nine
quartic couplings, making it hard to derive necessary and sufficient
BFB conditions in full generality. Therefore I considered the physically
well motivated approximation where one of the interactions is negligible,
in which case one can use well known results on the co-positivity
of matrices to derive the relevant conditions. For those cases where
all quartic couplings are relevant, I also derived a sufficient (but
not necessary) condition which can be used to pick acceptable coupling
constants.

\section*{Acknowledgments}

I acknowledge the financial support from MCIN/AEI (10.13039/501100011033)
through grant number PID2019-106087GB-C22, from the Junta de Andalucía
through grant number P18-FR-4314 (FEDER), from the Grant Agency of
the Czech Republic (GA\v{C}R) through contract number 20-17490S, and
also from the Charles University Research Center UNCE/SCI/013.

\section*{Appendix}

As discussed in the main text, the $2n$ vectors in equations (\ref{eq:vector_x})
and (\ref{eq:vector_y}) can be used to identify the allowed values
of the $\alpha$ and $\beta$ variables defined in expression (\ref{eq:alpha_beta}).
In particular, at the border of the valid $\alpha\beta$-region, these
vectors must either be null or point in a single direction. By carefully
considering the right-hand side of the expressions (\ref{eq:vector_x})
and (\ref{eq:vector_y}), one of the following possibilities must
be true.
\begin{enumerate}
\item For all $i$ such that $x_{i}\neq0$ (there must be at least one such
case since $\sum_{i}x_{i}=1$) we have $y_{i}=\beta$. We can further
divide this possibility in three cases.
\begin{enumerate}
\item $\beta=0$. This means that $x_{i}\neq0$ implies $y_{i}=0$. So we
can have at most $n-1$ non-zero $y_{i}$ which in turn means that
$\alpha\in\left[1/(n-1),1\right]$.
\item $\beta=\alpha\neq0$. In this scenario, the vectors (\ref{eq:vector_x})
and (\ref{eq:vector_y}) are aligned only if the value of all non-zero
$x_{i}$ or all non-zero $y_{i}$ is $\beta$($=\alpha$). So we conclude
that $\alpha=\beta=1/m$ where $m$ is the number of $x_{i}$ or $y_{i}\neq0$.
\item $\beta\neq0,\alpha$. This is undoubtedly \textbf{the most important
case}. By assumption, if $x_{i}\neq0$ then $y_{i}=\beta$, and for
all such cases the vectors $\left(2\left(\beta-\alpha\right),x_{i}-\beta\right)^{T}$
are proportional to each other only if the $x_{i}$ take a constant
value. In other words, there are $m$ non-zero $x_{i}$ and they all
have the same value $1/m$ (because $\sum_{i}x_{i}=1$), plus the
corresponding $y_{i}$ are equal to $\beta$. Any additional non-zero$y_{i}$
must be paired with a null $x_{i}$, and in all such cases the alignment
of the vectors $\left(2\left(y_{i}-\alpha\right),\beta\right)^{T}$
and $\left(2\left(\beta-\alpha\right),1/m-\beta\right)^{T}$ requires
that those $y_{i}$ also have a constant value given by the expression
$y_{i}=\left(m\beta^{2}-\alpha\right)/\left(m\beta-1\right)\equiv\omega$.
Let us assume that there are $m^{\prime}$ such occurrences; the overall
picture is this: there are $m$ cases $\left(x_{i},y_{i}\right)=\left(1/m,\beta\right)$,
$m^{\prime}$ occurrences of $\left(x_{i},y_{i}\right)=\left(0,\omega\right)$
and all other $\left(x_{i},y_{i}\right)$ are equal to $\left(0,0\right)$.
From the relation $1=\sum_{i}y_{i}=m\beta+m^{\prime}\omega$  we
conclude that
\begin{equation}
\alpha=\frac{1-2m\beta+\left(m^{2}+mm^{\prime}\right)\beta^{2}}{m^{\prime}}\,.
\end{equation}
The non-negative integers $m$ and $m^{\prime}$ can take any values
as long as $m\geq1$ and $m+m^{\prime}\leq n$. However, note that
the case $m=1$ and $m^{\prime}=n-1$ leads to the smallest value
of $\alpha$ (for any fixed $\beta$). This important setup corresponds
to the quadratic dependence of $\alpha$ on $\beta$ shown in equation
(\ref{eq:lower-part-of-figure}) which defines a line of utmost importance
for the extraction of the boundedness from below condition of the
scalar potential given in (\ref{eq:scalar-potential}).
\end{enumerate}
\item There is at least one $i$ such that $x_{i}\neq0$ and the corresponding
$y_{i}\neq\beta$. If this is the case, then all $y_{i}$ must either
be $0$ or $\alpha$ in order for the vectors $y_{i}\left(y_{i}-\alpha,x_{i}-\beta\right)^{T}$
to be collinear with $\left(0,1\right)^{T}$. If we were to call $m$
to the number of $y_{i}$ different from zero (this must be an integer
between 1 and $n-1$\footnote{At least one $y_{i}$ must be null. Otherwise, if $m=n$ then all
the $y_{i}$ would have the value $\alpha$ and it would follow that
$\beta=\sum_{i}x_{i}y_{i}=\alpha$, in contradiction with the assumption
that some $y_{i}\neq\beta$.}), then we conclude from $\sum_{i}y_{i}=1$ that $\alpha=1/m$. The
$x_{i}$ are unconstrained in this scenario, so it follows that $\beta$
can be anywhere in the range $\left[0,\alpha\right]$ (the value $\beta=0$
is reached for example when a single $x_{i}=1$ is paired with a null
$y_{i}$; on the other hand when all null $y_{i}$ have an associated
$x_{i}=0$ then $\beta=\alpha$).
\end{enumerate}
The four cases above (1.a, 1.b, 1.c and 2) correspond only to potential
fragments of the border of the $\alpha\beta$-region. In fact, some
of them are in the interior of this space. Figure \ref{fig:Allowed-ab}
depicts the actual border: the vertical line with $\beta=0$ and $\alpha\in\left[1/(n-1),1\right]$
(case 1.a), the horizontal line $\alpha=1$ and $\beta\in\left[0,1\right]$
(case 2 with $m=1$) and the parabola (\ref{eq:lower-part-of-figure})
with $\beta\in\left[0,1\right]$ (case 1.c).

Note that for a fixed value of $\alpha$, if we manage to find two
valid values of $\beta$ then all values in between them are equally
achievable.\footnote{One can see that it is so with the following reasoning. By definition
$\alpha\equiv\sum_{i}\left(y_{i}\right)^{2}=y^{T}y$ and $\beta=\sum_{i}x_{i}y_{i}=x^{T}y$
with the restriction that $\sum_{i}x_{i}=\sum_{i}y_{i}=1$ so we may
replace $x$ with $x^{\prime}=ty+\left(1-t\right)x$ where $t$ is
some number between 0 and 1. This replacement preserves $\alpha$
but changes $\beta$ to $\beta^{\prime}=\beta+t\left(\alpha-\beta\right)$,
which means that if a point $\left\{ \alpha,\beta\right\} $ is valid,
so is any point $\left\{ \alpha^{\prime},\beta^{\prime}\right\} $
in the line with $\alpha^{\prime}=\alpha$ and $\beta^{\prime}\in\left[\alpha,\beta\right]$
(or $\left[\beta,\alpha\right]$ if $\alpha>\beta$). As a consequence,
if two valid points have the same $\alpha$ and distinct $\beta$'s,
then all points in between them are equally allowed.} Using this fact, we conclude that all the space inside the border
(shaded area in figure \ref{fig:Allowed-ab}) is allowed as well.

\clearpage{}

\end{document}